\documentclass[12pt]{article}
\usepackage{putex}
\usepackage{graphicx}
\usepackage{amsmath}

\begin{document}

\preprint{PUPT-2247 \\ LMU-ASC 70/07}

\institution{PU}{Joseph Henry Laboratories, Princeton University, Princeton, NJ 08544, USA}
\institution{MaxPlanck}{Ludwig-Maximilians-Universit\"at, Department f\"ur Physik, Theresienstrasse 37, \cr 80333 M\"unchen, Germany}

\title{Shock waves from heavy-quark mesons in AdS/CFT}

\authors{Steven S. Gubser,\worksat{\PU,}\footnote{e-mail: {\tt ssgubser@Princeton.EDU}}
Silviu S. Pufu,\worksat{\PU,}\footnote{e-mail: {\tt spufu@Princeton.EDU}}
and Amos Yarom\worksat{\MaxPlanck,}\footnote{e-mail: {\tt yarom@theorie.physik.uni-muenchen.de}}}

\abstract{We calculate the far-field stress-energy tensor of a quark gluon plasma due to a heavy-quark meson moving through it, using a semi-classical description in string theory of the meson as a string hanging into anti-de Sitter space.  We find that these mesons create a shock wave but no diffusion wake, in contrast to single heavy quarks described via a trailing string, where the diffusion wake is strong.  If thermal QCD responds similarly to heavy quarks and heavy quarkonia, the presence or suppression of the diffusion wake constitutes a prediction which can in principle be checked in heavy-ion collisions.}

\date{November 2007}

\maketitle

\tableofcontents

\section{Introduction}

As a quark moves through a quark gluon plasma (QGP), it loses energy to the medium.  An extensive theoretical enterprise aims at describing this energy loss using some combination of perturbative QCD methods and hydrodynamics: for recent reviews see for example \cite{Majumder:2007iu,Casalderrey-Solana:2007km}.  There is also a substantial effort to compare energy loss in QCD to calculations in gauge theories with known holographic duals, especially ${\cal N}=4$ super-Yang-Mills: papers in this genre that directly set the stage for the current work are \cite{Herzog:2006gh,Casalderrey-Solana:2006rq,Gubser:2006bz,Friess:2006fk,Liu:2006nn,Chernicoff:2006hi,Gubser:2007xz,Chesler:2007an,Gubser:2007ga,Gubser:2007ni}, and related works include \cite{Liu:2006ug,Peeters:2006iu,Argyres:2006vs,Lin:2007pv}.  In \cite{Gubser:2007ni} it was shown that in a large class of theories, whose gravity duals may be described by asymptotically AdS black hole geometries, the ratio of energy from an infinitely heavy quark going into sound modes to the energy coming in from the diffusion wake behind the moving quark is $1+v^2:1$.  Thus the diffusion wake seems to be an important and persistent feature of the dynamics.  On the other hand, according to the analyses of \cite{CasalderreySolana:2006sq,Renk:2006mv}, PHENIX data supporting the hypothesis of jet splitting  \cite{Adler:2005ee} appears to favor a scenario where the wake is either absent or relatively suppressed.  It is worth noting that published results from STAR \cite{Adams:2005ph} with more inclusive transverse momentum cuts do not show jet splitting so much as substantial medium-induced jet broadening.

In this context it is interesting to inquire what pattern of energy flow surrounds other probes of ${\cal N}=4$ super-Yang-Mills theory.  Next to a single infinitely massive quark, described by a string with one end attached to the boundary of AdS-Schwarzschild and the other trailing behind it, the simplest such object is a heavy quarkonium-like state, described by a string with both ends attached to the boundary \cite{Maldacena:1998im}.  Such objects have been considered in \cite{Liu:2006nn,Chernicoff:2006hi}, where it was observed that they experience no drag force, provided their velocity is not too large relative to their size times the ambient temperature. One may speculate that the absence of a diffusion wake is generally associated with no-drag configurations, meaning configurations where the drag force is not visible in the approximation used to describe it (large $N$ in our case).\footnote{S.S.G.~thanks H.~Liu for a discussion that suggested this speculation.}

The aim of this paper is to compute the large distance asymptotics of the gauge theory stress-energy tensor produced by such heavy quarkonium states, both when the velocity is parallel to the separation of the quark and anti-quark, and when the velocity is perpendicular to it. We find that in both cases a sonic boom is observed but no diffusion wake. This may be compared with a phenomenological model \cite{Casalderrey-Solana:2004qm} used to explain the jet-splitting effect. There
it was assumed that when a single quark moves through the medium, the total energy loss is a subleading effect and the diffusion wake is suppressed.  Since \cite{Gubser:2007ga,Gubser:2007ni} show that this assumption does not hold for heavy quarks in  a large class of theories with asymptotically AdS duals, our result provides a first hint that jet-splitting (to the extent that it really occurs) may involve the motion of composite or colorless particles through the QGP.

There are two main steps in the computation of the gauge theory stress-energy tensor.  First, in section~\ref{S:Mesonic strings}, we review the string configuration of interest and identify its five-dimensional stress-energy tensor.  Second, in section~\ref{S:Metric}, we solve the linearized five-dimensional Einstein equations sourced by this five-dimensional stress-energy tensor.  Our solution is entirely analytic but is only accurate to leading order in small wave-numbers.  From it, the long-distance behavior of the gauge theory stress-energy tensor can be read off directly.  We continue in section~\ref{DISCUSSION} with some comparisons to hydrodynamics and a discussion of the maximum velocity for the mesons we are considering.

\section{Test string description of heavy-quark mesons}
\label{S:Mesonic strings}

The purpose of this section is to review the string configurations that describe the heavy-quark mesons of interest and to find out how these configurations source perturbations of the bulk metric.  The starting point is the Einstein-Hilbert plus Nambu-Goto action, conveniently expressed as
\begin{equation}
\label{ActionNeq4}
    S=\frac{1}{2\kappa_5^2}\int d^5x \left[\sqrt{-G}\,\left(R+{12\over L^2}\right)-\frac{2\kappa_5^2}{2\pi\alpha^{\prime}}
        \int d^2\sigma \sqrt{-g}\,\delta^5(x^\mu-X^{\mu})\right]\,.
\end{equation}
Here $G_{\mu\nu}$ is the space time metric, and $g_{ab}$ is the induced metric on the worldsheet.  The $X^\mu$ are the embedding functions for the string.
The background line element is
\begin{equation}
\label{E:AdSSSLine}
    ds^2 = \alpha(z)^2\left[-h(z) dt^2 + \sum_{i=1}^3 (dx^i)^2 + {dz^2\over h(z)} \right]
\end{equation}
where
 \eqn{alphaHDefs}{
  \alpha(z) = {L\over z} \qquad\hbox{and}\qquad
    h(z) = 1-\frac{z^4}{z_{\hbox{\tiny H}}^4}\,.
 }
This translation-invariant black hole background describes an infinite, static, thermal medium of ${\cal N}=4$ super-Yang-Mills.

For simplicity, we will only consider the cases where the quark-anti-quark separation is parallel to the direction of motion, or perpendicular to it.  In a gauge where $X^5=z$ and $X^0 = t$ we may use the ans\"atze
\begin{subequations}
\label{E:ansatz}
\begin{align}
\label{E:ansatzPll}
    X_{\parallel}^{\mu} &= \left(t, vt+\xi_{\parallel}(z),0,0,z\right)\\
\label{E:ansatzPpd}
    X_{\bot}^{\mu} &= \left(t, vt,\xi_{\bot}(z),0,z\right)
\end{align}
\end{subequations}
to describe the shape of the string \cite{Liu:2006nn,Chernicoff:2006hi,Peeters:2006iu,Argyres:2006vs}.  We shall usually drop the $\parallel$ and $\bot$ subscripts below, re-introducing them only when necessary.
One should be able to generalize the results here to the less symmetric case where the meson  separation is at an arbitrary angle $\theta$ relative to its velocity. It would also be interesting to consider extending this analysis to finite quark masses and to rotating mesons, along the lines of \cite{Peeters:2006iu}.

With the ansatz \eqref{E:ansatz}, $\xi$ is double valued, so we need to specify $\xi$ separately in the range $X^1-vt>0$ and $X^1-vt<0$ for the parallel case, and in the range $X^2>0$ and $X^2<0$ in the perpendicular case.
Consider first the solution at $X^1-v t>0$ (or $X^2 > 0$). In this region, we define $\xi_+ = \xi - \frac{1}{2}\ell$, $\ell$ being the distance between the quarks.  If $z_m$ is the maximum value of $z$ along the string, then the boundary conditions satisfied by $\xi_+$ are
 \eqn{E:BCxiplus}{
    \xi^{\prime}_+(z_m) &= -\infty\cr
    \xi_+(0) &= 0\,.
 }
Similarly, at $X^1-v t<0$ (or $X^2 < 0$) we define $\xi_- = \xi + \frac{1}{2}\ell$, which satisfies the boundary conditions
 \eqn{E:BCximinus}{
    \xi^{\prime}_-(z_m) &= \infty\cr
    \xi_-(0) &= 0\,.
 }
Putting these two solutions together, one finds a continuous, differentiable curve describing the shape of the string.

Because the Lagrangian is time independent, the momentum $\Pi_{\xi}$ conjugate to $\xi$ is conserved.  Explicitly, these momenta are
\begin{subequations}
\label{E:DefofPixi}
\begin{align}
\label{E:PixiPll}
    \Pi_{\xi_{\parallel}} &= -\frac{1}{2\pi\alpha^{\prime}}\frac{\xi_{\parallel}^{\prime} h\alpha^2}
        {\sqrt{1-\frac{v^2}{h}+\xi_{\parallel}^{\prime\,2}h}}\\
\label{E:PixiPpd}
    \Pi_{\xi_{\bot}} &= -\frac{1}{2\pi\alpha^{\prime}}\frac{\xi^{\prime}_{\bot} \left(h-v^2\right)\alpha^2}
        {\sqrt{1-\frac{v^2}{h}+\xi_{\bot}^{\prime\,2}\left(h-v^2\right)}}\,.
\end{align}
\end{subequations}
We define the quantities
\begin{subequations}
\label{E:DefofPi}
\begin{align}
    \Pi_{\parallel} &\equiv \frac{\sqrt{h(z_m)}\alpha(z_m)^2}{2\pi\alpha^{\prime}}\\
    \Pi_{\bot} &\equiv \frac{\sqrt{h(z_m)-v^2}\alpha(z_m)^2}{2\pi\alpha^{\prime}}\,,
\end{align}
\end{subequations}
which coincide with \eqref{E:DefofPixi} for $\xi = \xi_{+}$ after implementing the boundary condition $\xi^{\prime}_{+}(z_m) = -\infty$.
By inverting \eqref{E:DefofPixi}, $\xi_+$ can be written in integral form. For the case where the string is perpendicular to the direction of the velocity, this integral can be carried out explicitly in terms of Appell functions \cite{Friess:2006rk}.
From symmetry, we have $\xi_- = -\xi_+$. Also $\ell = -2\xi_+(z_m)$. Of the two possible configurations which solve $\ell = -2\xi_+(z_m)$ only the one in which the string is closer to the boundary is stable \cite{Friess:2006rk,Avramis:2006nv,Avramis:2007mv}. We shall work with this configuration.

By definition, the bulk stress-energy tensor tensor of the string is the quantity $J_{\mu\nu}$ that enters into the Einstein equations as
 \eqn{EinsteinJmunu}{
  R_{\mu\nu} - {1 \over 2} G_{\mu\nu} R -
    {6 \over L^2} G_{\mu\nu} = \kappa_5^2 J_{\mu\nu}\,.
 }
Starting from \eno{ActionNeq4}, one straightforwardly finds
\begin{equation}
\label{E:Jstring}
        J^{\mu\nu} = -\frac{1}{2\pi\alpha^{\prime}}
        \frac{\sqrt{-g}}{\sqrt{-G}}\partial X^{\mu} \partial X^{\nu}
        \delta(x^1-vt-X^1)\delta(x^2-X^2)\delta(x^3)\Theta(z_m-z)
\end{equation}
where
\begin{equation}
\label{E:DefofTheta}
 \Theta(x) = \begin{cases}
	1 & x>0 \\
	0 & x<0\,.
	\end{cases}
\end{equation}
This stress-energy tensor may conveniently be expressed in terms of a co-moving Fourier transform, meaning a three-dimensional Fourier transform in the $\vec{x}$ directions, with time dependence included by using $x^1-vt$ in place of $x^1$ in the exponent:
 \eqn{E:FourierTransform}{
  J^{\mu\nu}(t,\vec{x},z) =
    \int {d^3 k \over (2\pi)^3} J^{\mu\nu}(\vec{k},z)
      e^{i [k_1 (x^1-vt) + k_2 x^2 + k_3 x^3]} \,.
 }
Denoting
 \eqn{CSDef}{
    C_i &= \cos(k_i \left(\xi_++\ell/2\right))\cr
    S_i &= \sin(k_i \left(\xi_++\ell/2\right))\,,
 }
we find
\begin{multline}
\label{E:JPll}
    J^{\mu\nu}_{\parallel}(\vec{k}) = -
        \frac{\Pi_{\parallel}}{\alpha^7 h} \Theta(z_m-z) \times
    \\
    \begin{pmatrix}
        2 C_1 (1+h\xi_+^{\prime\,2})/h \xi_+^{\prime} &
        2 C_1 v/h \xi_+^{\prime} &
        0 & 0 &
        2 i S_1 v \\
        2 C_1 v/h \xi_+^{\prime} &
        -2 C_1 (-v^2+h^2 \xi_+^{\prime\,2})/h \xi_+^{\prime} &
        0 & 0 &
        2 i S_1 h \\
        0 &
        0 &
        0 & 0 &
        0 \\
        0 & 0 & 0 & 0 & 0 \\
        2 i S_1 v &
        2 i S_1 h &
        0 & 0 &
        2 C_1 (v^2 - h)/\xi_+^{\prime}
    \end{pmatrix}
\end{multline}
for the parallel  configuration
and
\begin{multline}
\label{E:JPpd}
    J^{\mu\nu}_{\bot}(\vec{k}) = -
        \frac{\Pi_{\bot}}{\alpha^7 (h-v^2)} \Theta(z_m-z) \times
    \\
    \begin{pmatrix}
        2 C_2 (1+h\xi_+^{\prime\,2})/h \xi_+^{\prime} &
        2 C_2 v (1+h \xi_+^{\prime\,2})/h \xi_+^{\prime} &
        0 & 0 &
        0\\
        2 C_2 v (1+h \xi_+^{\prime\,2})/h \xi_+^{\prime} &
        2 C_2 v^2 (1 + h \xi_+^{\prime\,2})/h \xi_+^{\prime} &
        0 & 0 &
        0 \\
        0 &
        0 &
        2 C_2 (v^2-h)\xi_+^{\prime} & 0 &
        -2 i S_2 (v^2 - h) \\
        0 & 0 & 0 & 0 & 0 \\
        0 &
        0 &
        -2 i S_2 (v^2 - h) & 0 &
        2 C_2 (v^2 - h)/\xi_+^{\prime}
    \end{pmatrix}
\end{multline}
for the perpendicular one.

\section{The gauge theory stress-energy tensor}
\label{S:Metric}

The computation of the gauge theory stress-energy tensor proceeds by solving the linearized Einstein equations sourced by \eno{E:JPll} or \eno{E:JPpd}.  This computation is non-trivial even in the leading long-distance approximation we will use in section~\ref{LONG}.  To set the stage, we review in section~\ref{GENERALSTRESS} the general approach for computing the holographic stress-energy tensor, and we discuss in section~\ref{SYMMETRIES} the symmetries of the problem as well as the counting of integration constants.

\subsection{Computing the holographic stress-energy tensor}
\label{GENERALSTRESS}

The metric perturbations sourced by the mesonic string are dual to fluctuations of the stress-energy tensor of the plasma according to the usual dictionary of AdS/CFT \cite{Gubser:1998bc,Witten:1998qj}.  Further developments, for example \cite{Balasubramanian:1999re,deHaro:2000xn,Bianchi:2001kw}, have clarified how to use this dictionary to perform calculations of appropriately regularized one-point functions of the type we are interested in.  The approach used here is essentially that of \cite{Friess:2006fk}, which gives the same results in this case as the more formal procedures of \cite{deHaro:2000xn,Bianchi:2001kw} once certain divergent contact terms, to be described below, are subtracted.
We start with a metric of the form
\begin{equation}
    G_{\mu\nu} = G_{\mu\nu}^{(0)}+\kappa_5^2 \alpha^2 H_{\mu\nu}\,,
\end{equation}
where $G_{\mu\nu}^{(0)}$ is the AdS${}_5$-Schwarzschild background \eno{E:AdSSSLine} and $H_{\mu\nu}$ is small compared to $G_{\mu\nu}^{(0)}$ (at least close to the conformal boundary.) The procedure is first to pass to axial gauge, $H_{\mu z} = 0$, and then to expand the remaining components near the conformal boundary:
 \eqn{BdyAsymp}{
  H_{mn} = P_{mn} z^3 + Q_{mn} z^4 + O(z^5)\,,
 }
where we have already used the boundary condition $H_{mn}(0)=0$ implying that the boundary theory metric is Minkowski. Without this boundary condition we would have gotten
\begin{equation}
\label{E:Deformation}
    H_{mn} = R_{mn} + S_{mn} z^2 + \mathcal{O}(z^3).
\end{equation}
The indices $m$ and $n$ run from $0$ through $3$ corresponding to the Minkowski space coordinates $(t, x^1, x^2, x^3)$.
With the boundary conditions leading to \eqref{BdyAsymp}, the leading behavior of $H_{mn}$ is $P_{mn} z^3$, but the coefficients $P_{mn}$ are found to be analytic in $\vec{k}$, corresponding to singularities with delta-function support in position space.  In principle, these singularities should contribute to the stress-energy tensor, and their contribution includes an infinite prefactor due to the scaling with $z^3$ rather than $z^4$.  Such divergent terms also appear when considering static quarks, and they are a result of the divergent mass of the quarks. In \cite{Maldacena:1998im} it was argued that these types of contributions should be dropped in order to obtain finite energy configurations.  As in \cite{Friess:2006fk}, we shall use a prescription where these terms are subtracted. A more detailed analysis along the lines of \cite{deHaro:2000xn,Bianchi:2001kw} should give identical results.
With this prescription, the expectation value of the gauge theory stress-energy tensor is
\eqn{TotalTmn}{
  \langle T_{mn} \rangle = \langle T_{mn} \rangle_{\rm bath} +
    \langle \delta T_{mn} \rangle \,,
 }
where
 \eqn{TmnBath}{
  \langle T_{mn} \rangle_{\rm bath} =
   {\pi^2 \over 8} N^2 T^4 \diag\{ 3,1,1,1 \}
 }
and
 \eqn{E:TBoundary}{
  \langle \delta T_{mn} \rangle = 2L^3 Q_{mn}
 }
is the contribution from the meson.  The coefficients $Q_{mn}$ are non-analytic in $\vec{k}$, so they capture long-distance behavior of $\langle \delta T_{mn} \rangle$.  As we will see, they signal the existence of a sonic boom but no diffusion wake.

\subsection{Symmetries and the counting of integration constants}
\label{SYMMETRIES}

In principle, the analysis ending in \eno{E:TBoundary} can be used to extract the holographic stress-energy tensor even for metrics that are finite deformations of the AdS${}_5$-Schwarzschild metric: all that matters is that the deformation should become small near the conformal boundary.  But it is very difficult to solve the full non-linear Einstein equations in the presence of a localized source like the strings described in section~\ref{S:Mesonic strings}.  We therefore pass to the linearized approximation of the Einstein equations:
 \eqn{LinearizedEinstein}{
  {\cal D}_{\mu\nu}^{\phantom{\mu\nu}\rho\sigma} H_{\rho\sigma} = J_{\mu\nu}\,,
 }
where ${\cal D}_{\mu\nu}^{\phantom{\mu\nu}\rho\sigma}$ is a variant of the Lichnerowicz operator.  The linearized approximation should be valid at large $N$, because in this approximation, the radius of AdS${}_5$ is much larger than the length scale in five dimensions where gravitational fields sourced by the string become strong.

Recall that we work throughout in axial gauge, where $H_{\mu z} = 0$.  Also, we use the co-moving ansatz for metric perturbations,
 \eqn{ComovingHmn}{
  H_{mn}(t,\vec{x},z) = \int {d^3 k \over (2\pi)^3}
    e^{i [k_1 (x^1-vt) + k_2 x^2 + k_3 x^3]} H_{mn}(\vec{k},z) \,.
 }
Plugging \eno{ComovingHmn} into \eno{LinearizedEinstein}, with either \eno{E:JPll} or~\eno{E:JPpd} as source terms, yields fifteen inhomogeneous differential equations for the co-moving Fourier coefficients $H_{mn}(\vec{k},z)$.  These are ordinary differential equations in the variable $z$.  An extended discussion of their structure and the appropriate boundary conditions
in the case of a cylindrically symmetric source (such as the inline meson)
was given in \cite{Friess:2006fk}.

The ten equations with $\mu$ and $\nu$ ranging from $0$ to $3$ (briefly, the $mn$ equations) are second order, so it takes twenty integration constants to specify a solution.  These twenty constants are the coefficients $R_{mn}$ and $Q_{mn}$ entering into \eno{BdyAsymp} and~\eno{E:Deformation},\footnote{The $P_{mn}$ are uniquely fixed by the behavior of the string near the boundary.} and setting $R_{mn}=0$ amounts to fixing ten of them.  The five equations with $\mu=z$ are first order constraints that fix five additional integration constants.  The constraint from the $zz$ equation is
\begin{equation}
\label{E:TracelessnessConstraint}
	Q_{00}-Q_{11}-Q_{22}-Q_{33} = 0\,,
\end{equation}
which is nothing but the condition for tracelessness of the stress-energy tensor of the $\mathcal{N}=4$ theory.
If we denote
 \eqn{kmDef}{
  k^m = \begin{pmatrix}
    v k_1 & k_1 & k_2 & k_3
  \end{pmatrix}\,,
 }
then the four remaining constraints, from the $mz$ equations, are
\begin{subequations}
\label{E:Constraints}
\begin{align}
\label{E:CPll}
	i k^m Q_{n m} &= i L^{-3} \Pi_{\parallel} \sin\left(\frac{k_1 \ell}{2}\right)
     \begin{pmatrix}
      v & -1 & 0 & 0
     \end{pmatrix} \\
\intertext{for the inline configuration, and}
\label{E:CPpl}
	i k^m Q_{n m} &= i L^{-3} \Pi_{\bot} \sin\left(\frac{k_2 \ell}{2}\right)
     \begin{pmatrix}
      0 & 0 & -1 & 0
     \end{pmatrix}
\end{align}
\end{subequations}
for the perpendicular configuration.  Equations \eqref{E:Constraints} represent the energy conservation equations in the boundary gauge theory. The remaining five integration constants are determined by the boundary conditions at the horizon which we discuss below.

\subsection{Solving the linearized Einstein equations for small $k$}
\label{LONG}

Having summarized the information that can be extracted from the first order constraint equations (the $\mu z$ equation in the axial gauge notation of the previous section), it is now time to attack the second order equations, coming from $mn$ components of the linearized Einstein equations \eno{LinearizedEinstein}.  It was shown in \cite{Friess:2006fk} how cylindrical symmetry around the direction of motion allows these equations to be partially decoupled; however, this method helps us directly only in the case where the quark separation is along the direction of motion.  The key simplification arises from isolating the leading small $k$ behavior.\footnote{Short distance (large $k$) asymptotics are likely to be obtained via the methods developed in \cite{Yarom:2007ap,Gubser:2007nd,Yarom:2007ni}.}  This behavior is expected to capture the large $x$ asymptotic behavior of $\langle \delta T_{mn} \rangle$ because the propagating modes are hydrodynamic: non-hydrodynamical modes have some attenuation length and therefore decay exponentially at large distances from the source. The simplification at small $k$ is that the $mn$ equations naturally decouple and take the form
\begin{equation}
\label{E:EOM}
    \alpha^{-3} h^{-n}\frac{d}{dz}\left(X' \alpha^3 h^n\right) + \mathcal{O}(k^2) = S_X\,,
\end{equation}
where $X$ corresponds to various linear combinations of $H_{mn}$ and $S_X$ is the corresponding combination of $J_{mn}$'s.\footnote{With a slight abuse of notation, we shall only include the indices of $X$ in the subscript of $S_X$. Thus, in \eqref{E:DefS0i} and \eqref{E:Sij}, instead of $S_{H_{mn}}$ we shall write $S_{mn}$. Similarly, in \eqref{FDefs} and \eqref{SDefs} we use $S_i$ instead of $S_{F_i}$ .} More explicitly, we find
that
$H_{01},\,H_{02}$ and $H_{03}$ satisfy \eqref{E:EOM} with $n=0$ and
\begin{equation}
\label{E:DefS0i}
    S_{0i} = -{2 \over h} J_{0i}\,.
\end{equation}
The other off-diagonal components $H_{12}$, $H_{13}$ and $H_{23}$ satisfy \eqref{E:EOM} with $n=1$ and
\begin{equation}
\label{E:Sij}
	S_{ij} = - {2 \over h} J_{ij}\,.
\end{equation}
The diagonal components need to be reshuffled in order to be brought in the form \eqref{E:EOM}.  Defining
 \eqn{FDefs}{
    F_0 &= H_{11}+H_{22}+H_{33}\cr
    F_1 &= -\frac{2}{3}H_{11}+\frac{1}{3}H_{22}+\frac{1}{3}H_{33}\cr
    F_2 &= -H_{22}+H_{33}\cr
    F_3 &= -\frac{3 H_{00}}{h}+H_{11}+H_{22}+H_{33}
 }
we find that each of the $F_i$ satisfies equation \eqref{E:EOM} with $n_0 = 1/2$, $n_1 = n_2=1$ and $n_3 = 3/2$, and
 \eqn{SDefs}{
    S_0 &= -{2\over h^2} J_{00}\cr
    S_1 &= -\frac{2}{3h} \left(-2J_{11}+J_{22}+J_{33}\right)\cr
    S_2 &= -\frac{2}{h} \left(J_{33}-J_{22}\right)\cr
    S_3 &= \frac{2}{h} \left({1\over h} J_{00}+J_{11}+J_{22}+J_{33}\right)\,.
 }

The solution to equation \eqref{E:EOM} is
\begin{equation}
    \alpha(z)^{3} h(z)^{n} X^{\prime}(z) = q_X + \begin{cases}
        \int_{z_m}^z dx \, h(x)^{n} \alpha(x)^3 S_X(x)  & z<z_m \\
        0 & z>z_m \,,
    \end{cases}
\end{equation}
where the two branches of the solution arise from the step function in \eqref{E:Jstring}.
Thus,
\begin{equation}
\label{E:SolX}
    X(z) = r_X + \int_0^z dy \frac{q_X}{\alpha(y)^{3} h(y)^{n}} +
	\begin{cases}
        \int_0^z dy \frac{\int_{z_m}^y dx \, h(x)^{n} \alpha(x)^3 S_X(x)}{\alpha(y)^{3} h(y)^{n} } & z<z_m \\[1.5\jot]
        \int_0^{z_m} dy \frac{\int_{z_m}^y dx \, h(x)^{n} \alpha(x)^3 S_X(x)}{\alpha(y)^{3} h(y)^{n}} & z>z_m \,,
    \end{cases}
\end{equation}
where $q_X$ and $r_X$ are integration constants.  We will find it convenient to define
\begin{equation}
\label{E:DefSigma}
	\Sigma_X(z) = \int_{z_m}^z dx \, h(x)^{n} \alpha(x)^3 S_X(x)\,.
\end{equation}
Comparing \eno{E:SolX} to~\eno{BdyAsymp}, it is clear that the $r_X$ are related to $R_{mn}$, and the $q_X$ are related to the $Q_{mn}$.  So we should set $r_X=0$ for all ten choices of $X$ in order not to introduce a deformation of the metric in the boundary theory.

Even before working out the precise relationship between the $q_X$ and the $Q_{mn}$ (which follows from asymptotic expansions in small $z$ and use of \eno{FDefs}), we reason that the constraints \eqref{E:TracelessnessConstraint} and \eno{E:Constraints} must translate into five constraints among the $q_X$.  In order to uniquely specify the physical solution to the linearized Einstein equations, five more boundary conditions are needed.  As in the case of the trailing string, these come from horizon boundary conditions, which demand that there should be no outgoing modes at $z=z_{\hbox{\tiny H}}$.  Let's focus (slightly presciently) on the five choices of $X$ where $n=1$.  Then we see from \eno{E:SolX} that, near $z=z_{\hbox{\tiny H}}$,
 \eqn{XHor}{
	X(z) &= -\frac{z_{\hbox{\tiny H}}^4 q_X}{4 L^3} \ln h(z) +
		\int_0^{z_m} \frac{\Sigma_X(y)}{\alpha(y)^3 h(y)}dy \cr
	     & \sim q_X \log (z_{\hbox{\tiny H}}-z)+\ldots \,.
 }
The claim is that these $q_X$ (i.e.~the five corresponding to the choices of $X$ where $n=1$) have to be zero.  To see this, we reason that at finite $k$, outgoing modes have the form $X(z) \sim (z_{\hbox{\tiny H}} - z)^{i v k_1 z_{\hbox{\tiny H}}/4}$ near $z=z_{\hbox{\tiny H}}$, while infalling ones behave as $X(z) \sim (z_{\hbox{\tiny H}} - z)^{-i v k_1 z_{\hbox{\tiny H}}/4}$.  At finite $k$, we would allow the infalling solution and disallow the outgoing one.  In the limit of small $k$, one may expand
 \eqn{HorizonExpand}{
  (z_{\hbox{\tiny H}}-z)^{\pm i v k_1 z_{\hbox{\tiny H}}/4} \approx
    1 \pm {i v k_1 z_{\hbox{\tiny H}} \over 4} \log
      (z_{\hbox{\tiny H}}-z) \,.
 }
The coefficient of the log term in \eno{HorizonExpand} is suppressed by a power of $k$, hence the claim that $q_X=0$ in \eno{XHor}.\footnote{A finer analysis shows that $q_{13} = q_{23} = q_{12}=\mathcal{O}(k^2)$ while $q_1 = \mathcal{O}(k)$ and $q_2 = \mathcal{O}(k^2)$ for the inline meson and $q_2 = \mathcal{O}(k)$ for the perpendicular one.}  What is perhaps surprising about the argument following \eno{HorizonExpand} is that, at leading order in $k$, it doesn't distinguish between infalling and outgoing modes.  But that is in fact appropriate: dissipative effects, such as shear viscosity, are invisible at this order in $k$, so it's right to demand that there is neither inward nor outward flux at the horizon.  When a similar analysis is carried out for the three choices of $X$ where $n=0$ and the ones with $n=1/2$ and $n=3/2$, the outcome is that there is no constraint from the horizon on the corresponding $q_X$.

Having argued that all ten $r_X$ and five of the $q_X$ must equal zero, then together with the five constraints \eqref{E:TracelessnessConstraint} and \eqref{E:Constraints} we can determine all the $Q_{mn}$.
For the choices of $X$ where $n=1$ we write
 \eqn{XBdy}{
  X(z) = P_X z^3 + Q_X z^4 + {\cal O}(z^5)\,,
 }
where $P_X$ is related to the $P_{mn}$ terms in \eqref{BdyAsymp} and $Q_X$ to $Q_{mn}$.
Since $\Sigma_X(z) = \mathcal{O}(z^{-1})$, it follows from \eqref{E:SolX} that
 \eqn{E:Qq}{
  Q_X = {1\over 4 L^3} \lim_{z\to 0} \partial_z \left(z \Sigma_X(z) \right)\,.
 }
For both the inline and perpendicular configurations, the above equation combined with \eqref{E:DefSigma} and the corresponding source terms of the $Q_{12}$, $Q_{23}$, and $Q_{13}$ equations give
\begin{equation}
\label{E:QisZero}
	Q_{12}=Q_{23}=Q_{13}={\cal O}(k^2)\,.
\end{equation}
Thus far, the analysis is essentially the same for the perpendicular and inline cases.  In the remainder of the analysis we treat the perpendicular and inline cases separately.

For the perpendicular case, \eqref{E:DefSigma} combined with the appropriate source terms of the $F_1$ and $F_2$ equations gives
 \eqn{GotSigmaPpd}{
	\Sigma_1(z) &= -\frac{4}{3} \Pi_\perp \left(\xi_+(z)+\ell/2 - v^2 \int_{zm}^z dx \, \frac{2(1+h \xi_+^{\prime\,2})}{(v^2-h)h\xi_+^{\prime}} \right)\cr
	\Sigma_2(z) &= 4 \Pi_\perp \left(\xi_+(z)+\ell/2\right)
 }
where we have used $\xi_+(z_m) = -\ell/2$. The integral in \eqref{GotSigmaPpd} and $\xi_+(z)$ may be written in terms of Appell hypergeometric functions. From \eqref{E:Qq} and \eqref{GotSigmaPpd} we find
 \eqn{GotQPpd}{
	Q_1 &= -\frac{1}{6 L^3} \Pi_\perp \left(\ell - 4 v^2 \sigma_{\bot}\right)\cr
	Q_2 &= \frac{1}{2 L^3}\Pi_\perp \ell
 }
where
\begin{equation}
	\sigma_{\bot}(v,z_m) = \frac{2^{1/2}\pi^{3/2} z_m\, {}_2F_1\left(-\frac{1}{4},\frac{1}{2},\frac{1}{4},\frac{z_m^4}{z_{\hbox{\tiny H}}^4}\right)}{\Gamma\left(\frac{1}{4}\right)^2\sqrt{1-v^2}\sqrt{1-v^2-z_m^4/z_{\hbox{\tiny H}}^4}} \,.
\end{equation}
At this point, we have determined five of the $Q_{mn}$ for the perpendicular case.  The rest follow from conservation and tracelessness of the stress-energy tensor (or, in gravitational terms, the first order constraints).  Explicitly, by inverting the relations \eqref{FDefs} and using \eqref{E:TracelessnessConstraint},  \eqref{E:CPpl}, \eqref{E:QisZero} and \eqref{GotQPpd} we can obtain the stress-energy tensor through \eqref{E:TBoundary}, and it reads as follows:
\begin{equation}
\label{E:StressEnergyPpd}
	\langle \delta T_{\bot mn} \rangle = \frac{\Pi_{\bot}}{k^2 - 3 k_1^2 v^2} \left(\ell \tau^{(\bot)}_{mn} + v \sigma_{\bot} \tau^{(\sigma)}_{mn}\right) + {\cal O}(k)\,,
\end{equation}
with
\begin{equation}
\label{E:TauPpd}
	\tau^{(\bot)}_{mn} = \begin{pmatrix}
 -k^2 & k_1^2 v & k_1 k_2 v & k_1 k_3 v \\
 k_1^2 v & -k_1^2 v^2 & 0 & 0 \\
 k_1 k_2 v & 0 & 2 v^2 k_1^2-k^2 & 0 \\
 k_1 k_3 v & 0 & 0 & -k_1^2 v^2
	\end{pmatrix}
\end{equation}
and
\begin{equation}
\label{E:Tausigma}
	\tau^{(\sigma)}_{mn} = \begin{pmatrix}
 v \left(3 k_1^2-k^2\right) & k_{\bot}^2-2 k_1^2 v^2 & - k_1 k_2 \left(1- v^2\right) & - k_1 k_3 \left(1 - v^2\right) \\
 k_{\bot}^2-2 k_1^2 v^2 & -v \left(k_{\bot}^2-2 k_1^2 v^2\right) & 0 & 0 \\
 - k_1 k_2 \left(1-v^2\right) & 0 & k_1^2 v \left(1 - v^2\right) & 0 \\
 - k_1 k_3 v \left(1-v^2\right) & 0 & 0 & k_1^2 v \left(1 - v^2\right)
	\end{pmatrix}\,.
\end{equation}

For the inline case, we find that $\Sigma_{2} = \mathcal{O}(k)$ so that $Q_2 = \mathcal{O}(k)$, but
\begin{equation}
\label{E:Sigma1Pll}
	\Sigma_{1}(z) = \frac{8}{3} \Pi_\parallel \left(\xi_{+}(z)+\ell/2- v^2 \int_{z_m}^z dx\, \frac{1}{h^2(x)\xi^{\prime}_+(x)}\right)\,.
\end{equation}
Here we cannot carry out the second integral explicitly.
From \eqref{E:Qq} it follows that
\begin{equation}
	Q_{1} = \frac{\Pi_\parallel}{3 L^3}\left(\ell + 2 v^2 \sigma_{\parallel}\right)
\end{equation}
with
 \eqn{GotsigmaPllVar}{
	\sigma_{\parallel}(v,z_m) = \lim_{z \to 0} {\partial\over \partial z} \left[ z \int_z^{z_m} dx \frac{1}{h(x)^2 \xi_+^{\prime}(x)} \right] \,
 }
which can be written as a definite integral and carried out numerically, if desired, for specified $v$ and $z_m$.
Thus,
\begin{equation}
\label{E:StressEnergyPll}
	\langle \delta T_{\parallel\,mn} \rangle = \frac{\Pi_{\parallel}}{k^2 - 3 k_1^2 v^2} \left(\ell \tau^{(\parallel)}_{mn} + 2 v \sigma_{\parallel} \tau^{(\sigma)}_{mn}\right)
\end{equation}
where
\begin{equation}
\label{E:TauPll}
	\tau^{(\parallel)}_{mn} = \begin{pmatrix}
 -3 v^2 k_1^2-k^2 & 2 k_1^2 v & 2 k_1 k_2 v & 2 k_1 k_3 v \\
 2 k_1^2 v & v^2 k_1^2-k^2 & 0 & 0 \\
 2 k_1 k_2 v & 0 & -2 k_1^2 v^2 & 0 \\
 2 k_1 k_3 v & 0 & 0 & -2 k_1^2 v^2
	\end{pmatrix}
\end{equation}
and $\tau^{(\sigma)}_{mn}$ is the same as in \eqref{E:Tausigma}.  A mild consistency check of these results is that in the $v \to 0$ limit, $T_{\bot mn}$ coincides with $T_{\parallel mn}$ up to a ninety-degree rotation.

Equations \eno{E:StressEnergyPpd}--\eno{E:Tausigma} and \eno{E:StressEnergyPll}--\eno{E:TauPll} are our final results from solving the linearized Einstein equations in the presence of the mesonic string configurations, at leading order in small $k$. The two poles, located along the real $k_1$ axis at $k_1^2 = k_{\bot}^2/(3 v^2-1)$ for $v^2>1/3$ indicate that a shock wave will appear in real space. The absence of a pole at $k_1=0$ indicates that there is no diffusion wake \cite{Gubser:2007xz}.

\section{Discussion}
\label{DISCUSSION}

Calculations in AdS/CFT at small wave-number may reasonably be expected to relate to hydrodynamics.  This theme has been emphasized, for example, in the recent review \cite{Son:2007vk}; see also \cite{Bhattacharyya:2007vs} and references therein.  We pursue this comparison in section~\ref{HYDRO}.  In section~\ref{RESTRICTIONS} we remark briefly on the well-known speed limit for finite-mass mesons constructed as hanging strings and make some tentative remarks about the possible phenomenological relevance of our calculations.

\subsection{Comparison to hydrodynamics}
\label{HYDRO}

From \eqref{E:Constraints} it is clear that the source for the mesonic configuration has a dipole structure. In real space (at $t=0$) equations \eqref{E:Constraints} read
\begin{subequations}
\label{E:Conservation}
\begin{align}
	\partial^{m} T_{\parallel mn} = \left(v,-1,0,0\right)_{n}\Pi_{\parallel} \left(\delta(x_1-\ell/2) - \delta(x_1+\ell/2)\right)\delta(x_2)\delta(x_3)\\
	\partial^{m} T_{\bot mn} = \left(0,0,-1,0\right)_{n}\Pi_{\bot} \delta(x_1) \left(\delta(x_2-\ell/2) - \delta(x_2+\ell/2)\right)\delta(x_3)\,.
\end{align}
\end{subequations}
Let us compare \eqref{E:Conservation} to a hydrodynamic analysis. In the linearized hydrodynamic approximation, the stress-energy tensor may be written in terms of the velocity field. This amounts to six relations
\begin{equation}
\label{E:Constitutive}
	T^{ij}_{\rm hydro}  = c_s^2 \delta^{ij} T^{00}_{\rm hydro} - {3\over 4} \Gamma_s \left(\partial^i T^{0j}_{\rm hydro} + \partial^j T^{0i}_{\rm hydro} - \frac{2}{3} \delta^{ij} \partial_l T^{0l}_{\rm hydro}\right)
\end{equation}
between the space-space components of the stress-energy tensor, the energy density, and the Poynting vector.
The four additional relations needed to completely specify the stress-energy tensor are
\begin{equation}
\label{E:HydroConservation}
	i k_m T^{mn}_{\rm hydro} = J^{n}_{\rm hydro}\,.
\end{equation}
Naively one would like to identify $J^n_{\rm hydro}$ with the source terms in \eqref{E:Conservation}. However, in \cite{Casalderrey-Solana:2004qm,Gubser:2007ni} it was argued that, in the case of a moving quark, the quark together with its near field source the hydrodynamic modes, as opposed to just the quark itself.

In \cite{Gubser:2007ni} a (non-unique) prescription was used to compare the full stress-energy tensor, in our case  \eqref{E:StressEnergyPpd} and \eqref{E:StressEnergyPll}, to the one expected based on a hydrodynamic analysis:
by identifying
\begin{equation}\label{E:IdentifyHydro}
	\langle \delta T^{m0} \rangle = T_{\rm hydro}^{m0}\,,
\end{equation}
and using \eqref{E:Constitutive}
we find
\begin{subequations}
\label{E:Nonhydro}
\begin{align}
	\langle \delta T_{\parallel} \rangle &= T_{\parallel {\rm hydro}} + \frac{1}{3}\Pi_{\parallel} \left(\ell+2 v^2 \sigma\right)\diag\begin{pmatrix} 0 & -2 & 1 & 1 \end{pmatrix}\\
	\langle \delta T_{\bot} \rangle &= T_{\bot {\rm hydro}}
	+ \frac{1}{3}\Pi_{\bot} v^2 \sigma \diag\begin{pmatrix} 0 & -2 & 1 & 1 \end{pmatrix}
	+ \frac{1}{3}\Pi_{\bot} \ell \diag\begin{pmatrix} 0 & 1 & -2 & 1 \end{pmatrix}\,.
\end{align}
\end{subequations}
With these definitions, the hydrodynamic sources are given by
 \eqn{HydroSources}{
	J_{\rm hydro}^n =- \Pi \begin{pmatrix}
		i k_1 v \ell \delta &
		\frac{1}{3}i k_1 (\ell-4 v^2 \sigma) &
		\frac{1}{3}i k_2 (\ell+2 v^2 \sigma) &
		\frac{1}{3}i k_3 (\ell+2 v^2 \sigma)
		\end{pmatrix}
 }
where $\delta = 1$ for the parallel configuration and zero for the perpendicular one.
Adopting the notation of \cite{Casalderrey-Solana:2004qm} where
\begin{equation}
\label{E:CSTSource}
	J_n = \begin{pmatrix} e_0 & g_0 + k_1 g_1 & k_2 g_1 & k_3 g_1 \end{pmatrix}
\end{equation}
we find
\begin{subequations}
\label{E:e0g0g1}
\begin{align}
\label{E:e0}
	e_{\parallel\,0} &=  i k_1 \Pi v \ell &e_{\bot\,0} &=  0\\
\label{E:g0g1}
	g_{0} &=  2 i k_1 \Pi v^2 \sigma & g_{1} &=  - \frac{1}{3}\Pi \left(\ell + 2 v^2 \sigma\right)\,.
\end{align}
\end{subequations}

The source \eqref{E:CSTSource} was shown to produce a diffusion wake when $g_1$ vanishes, and following a Cooper-Frye treatment of hadronization the authors of \cite{Casalderrey-Solana:2004qm} concluded that no jet-splitting was predicted.  A similar analysis with $g_0=e_0=0$ led to a prediction of jet-splitting, although subsequent work \cite{Chaudhuri:2005vc,CasalderreySolana:2006sq} suggests that the rate of parton energy loss may have to be tuned to unrealistically high values in order to match to data.  In \cite{Gubser:2007ni} a figure of merit $\gamma_1$ was proposed to distinguish quantitatively between the two scenarios:
\begin{equation}
\label{E:gamma1}
	\gamma_1 = \frac{1}{R}\left|\frac{g_1(k=1/R)}{e_0(k=1/R)} \right|.
\end{equation}
The expectation is that large $\gamma_1$ leads to a prediction of jet-splitting, while small $\gamma_1$ does not.
Since $e_{\bot\,0} = 0$ we find immediately that $\gamma_{1\,\bot} \to \infty$,  though perhaps a more conservative approach would be to redefine
\begin{equation}
\label{E:gamma2}
	\gamma_1 = \frac{1}{R}\left|\frac{g_1(k=1/R)}{g_0(k=1/R)} \right|,
\end{equation}
which gives
\begin{equation}
	\gamma_1 = \frac{1}{3}\left(1 + \frac{\ell}{2 v^2 \sigma}\right).
\end{equation}

Caution is in order when using \eno{E:e0g0g1}, \eno{E:gamma1}, or \eno{E:gamma2}, because it's not clear that the quantities $e_0$, $g_0$, and $g_1$ can be defined in a way that is independent of contact terms, meaning terms which are analytic in $k$.  The reason for this is that while $\langle \delta T_{mn} \rangle$ has non-analytic behavior in $k$ and can therefore be understood as encoding universal behavior about the long-distance asymptotics, this non-analyticity vanishes when we use \eno{E:HydroConservation} and \eno{E:IdentifyHydro}, and some prescription for the treatment of contact terms is implied.  Despite these intricacies, the main qualitative features of our analysis is easy to state: for the heavy meson there is a sonic boom and no diffusion wake, and it's because of a dipole structure of the source.

\subsection{Restrictions on the velocity}
\label{RESTRICTIONS}

It is tantalizing to have a microscopic description---a rather simple one at that---of a source that avoids producing a diffusion wake.  But is this source just a contrivance, or does it have some relevance to QGP physics?  The charm quark is heavy enough compared to the temperature scale at RHIC for the calculations we have presented to be compared meaningfully to the behavior of the $J/\psi$ meson moving in a quark-gluon plasma.  The $J/\psi$ is indeed produced in heavy-ion collisions at RHIC, and some of its properties are studied: see for example \cite{PereiraDaCosta:2005xz} as well as results from CERN experiments at lower energies, e.g.~\cite{Alessandro:2004ap}. If one could construct a di-hadron correlator where one hadron is tagged as coming from a $J/\psi$, it would be reasonable from our analysis to expect a split-jet structure, corresponding to a suppressed diffusion wake.  But $J/\psi$ production is so rare at RHIC that it can't possibly be relevant to the medium modification of the away-side jet as seen in untagged di-hadron correlators, studied for example in \cite{Adams:2005ph,Adler:2005ee}.  In these studies, the overwhelming majority of energetic partons are probably energetic light quarks and gluons---or at least, states unrelated to heavy flavor. But ``light'' at RHIC energies probably should be understood to include the strange quark.  It is tempting then to speculate that strange quark mesons, for example the $\phi$, might exhibit the effect we describe, namely production of a sonic boom but not a diffusion wake (or a suppressed diffusion wake).  Even if it seems improbable that such an effect could explain jet-splitting as observed in \cite{Adler:2005ee}, one might reasonably inquire whether strange quark mesons are enhanced in a directional fashion in the final state.

One immediate question that comes to mind when considering this scenario is how the $\phi$ forms soon enough to have time to propagate through a significant amount of quark-gluon plasma. With our current techniques, our AdS/CFT analysis does shed light on this problem. What we do note, is that there is an upper bound on the energy of the meson which follows from a bound on its velocity;
finite mass quarks described by strings ending on branes (and mesons made from such quarks) cannot move faster than a certain velocity, calculated from the condition that the the endpoints of the string cannot move faster than light.  This observation seems to have been made in various nearly equivalent ways by several authors, for example \cite{Peeters:2006iu,Liu:2006nn,Chernicoff:2006hi,Gubser:2006nz}.  Here we will follow approximately the discussion of \cite{Gubser:2006nz}, adapted to the case of mesons.

The setup for finite-mass quarks involves a D-brane located at a finite value $z=z_*$ in the AdS${}_5$-Schwarzschild background, our equation \eno{E:AdSSSLine}.\footnote{Most simply, this can be a D7-brane, as in \cite{Karch:2002sh}.  Note that in this case, what $z_*$ means is the deepest inside AdS${}_5$ that the D7-brane reaches.}  The thermal mass of a static quark, first calculated in \cite{Herzog:2006gh}, is
 \eqn{mStatic}{
  m_{\rm quark} = {L^2 \over 2\pi\alpha'} \left(
    {1 \over z_*} - {1 \over z_{\hbox{\tiny H}}} \right) \,.
 }
An endpoint of the string moving at constant depth $z_*$ with a fixed coordinate velocity, say $x^1 = vt$, has an induced line element
 \eqn{InducedLine}{
  ds^2_{\rm induced} =
    \alpha^2 \left[ -h(z_*) + v^2 \right] dt^2 \,.
 }
Evidently, this induced line element is timelike when $v < v_*$, null when $v = v_*$, and spacelike when $v > v_*$, where
 \eqn{vstarDef}{
  v_* = \sqrt{h(z_*)} \,.
 }
Thus we conclude that $v_*$ is the maximum velocity that the meson is allowed to have.  The associated maximum Lorentz gamma factor has a simple form:
 \eqn{gammastarDef}{
  \gamma_* = {1 \over \sqrt{1-v_*^2}} = {z_*^2 \over z_{\hbox{\tiny H}}^2} =
   \left( 1 + {2\pi\alpha' \over L^2} m_{\rm quark} z_{\hbox{\tiny H}}
    \right)^2 =
   \left( 1 + {2m_{\rm quark} \over T \sqrt{g_{YM}^2 N}}
     \right)^2 \,,
 }
where in the last equality we have remembered that $T = 1/\pi z_{\hbox{\tiny H}}$ and $g_{YM}^2 N = L^2/\alpha'$.  Without attempting to account for the possibility of non-trivial dispersion relations, the maximum energy of a meson built as a string with both ends on the D-brane at $z=z_*$ is
 \eqn{Emax}{
  E_{\rm max} = \gamma_* m_{\rm meson} =
    m_{\rm meson} \left( 1 + {2m_{\rm quark} \over T
      \sqrt{g_{YM}^2 N}} \right)^2 \,.
 }

As a simple-minded example, imagine the $\phi$ meson, with $m_\phi \approx 1\,{\rm GeV}$, to be composed of $s$ and $\bar{s}$ with $m_s \approx 300\,{\rm MeV}$.  Suppose this meson propagates through a plasma with $T=250\,{\rm MeV}$ and $g_{YM}^2 N = 10$.  Then $E_{\rm max} \approx 3\,{\rm GeV}$, and $v_* \approx 0.95$.  While $v_*$ is comfortably above the speed of sound, $E_{\rm max}$ is slightly below the energy range in which there is some experimental support for the phenomenon of away-side ``punch-through'' (see for example \cite{Jia:2007sf}).\footnote{It should be noted that the peak at $\Delta\phi = \pi$ for high $p_T$ particles is not unambiguous evidence for high-energy particles going all the way through the QGP: an alternative scenario hinging on surface bias may also explain the data.  S.~Gubser thanks P.~Steinberg for a discussion on this point.}
One should keep in mind that in this example we've been applying an $\mathcal{N}=4$ result to QCD, so that a precise bound on the meson mass cannot be obtained. It should be clear though, that decreasing the quark mass in \eno{Emax} brings $E_{\rm max}$ down quickly, especially if $m_{\rm meson}$ is also decreased.
Thus, a heavy-light meson, which would be described in this context by a straight string stretching between two D7-branes \cite{Paredes:2004is,Herzog:2006gh,Erdmenger:2007vj}, is also problematic.

Our bottom line is that, while single heavy quarks as described in AdS/CFT generate a strong diffusion wake in addition to a sonic boom, heavy quarkonia in the same description generate only a sonic boom.  Applying these heavy quarkonium results to the $\phi$ meson is risky.  Even so, it would be interesting to inquire whether there is any directed excess of strange mesons associated with hard probes.  If such an excess exists, how does it relate to jet-splitting?

\section*{Acknowledgments}

A.~Y.~would like to thank J.~Casalderrey-Solana and E.~Shuryak for stimulating discussions.  The work of S.~Gubser was supported in part by the Department of Energy under Grant No.\ DE-FG02-91ER40671 and by the NSF under award number PHY-0652782.  A.~Yarom is supported in part by the German Science Foundation and by the Minerva foundation.

\clearpage
\bibliographystyle{ssg}
\bibliography{qqbar}

\end{document}